\documentclass[global,twocolumn]{svjour}
\usepackage{graphics}
\usepackage{amsmath}
\usepackage{braket}

\journalname{myjournal}
\begin{document}
\title{A trapped-ion local field probe}
\author{Gerhard Huber \and Frank Ziesel \and Ulrich Poschinger \and Kilian Singer \and Ferdinand Schmidt-Kaler
}                     
%
%
\institute{Institut f\"ur Quanteninformationsverarbeitung, Universit\"at Ulm, Albert-Einstein-Allee 11, 89069 Ulm, Germany}
\date{Received: date / Revised version: date}
%
\maketitle
\begin{abstract}
We introduce a measurement scheme that utilizes a single ion as a local field probe. The ion is confined in a segmented Paul trap and shuttled around to reach different probing sites. By the use of a single atom probe, it becomes possible characterizing fields with spatial resolution of a few nm within an extensive region of millimeters. We demonstrate the scheme by accurately investigating the electric fields providing the confinement for the ion. For this we present all theoretical and practical methods necessary to generate these potentials. We find sub-percent agreement between measured and calculated electric field values.
\end{abstract}
\section{Introduction}
\label{intro}
The concept to realize and investigate quantum processes with single ions in Paul traps was subject to an impressive development over the last years. As for quantum information with single to a few ions, there has hardly been an experimental challenge that could withstand this development, consider for instance high fidelity quantum gates~\cite{Benhelm}, multi-qubit entanglement~\cite{Haeffner,Leibfried}, and error correction~\cite{Chiaverini}. There has been a recent development in ion trap technology that was primarily initiated by the necessity -- and possibility -- to scale up the established techniques to process larger numbers of ions. This idea how to scale up an ion trap computer is to create multiple trap potentials, each keeping only a small, processable number of ions~\cite{Kielpinski}. Time dependent potentials can then be used to shuttle the ions between different sections. To generate such spatially separated potentials, the respective trap electrodes are segmented along the transport direction and supplied individually with time dependent voltages. There have also been proposals to incorporate the shuttling process into the quantum gate time by letting the ion cross laser beams or magnetic field arrangements~\cite{Leibfried2}. Another application of the shuttling ability is the exact positioning of the ion in the mode volume of an optical cavity to perform cavity quantum electrodynamic experiments. Recently, multi-trap configurations with locally controlled, individual trap frequencies were proposed to generate cluster states in linear ion traps to perform one-way quantum computations~\cite{Wunderlich}.

This paper presents the implementation of a method using a single trapped ion to investigate local electric fields. The method profits from the outstanding accuracy of spectroscopic frequency measurements and relies on the ability to shuttle the ion in order to reach different probing sites. First, we describe the experimental system and the way the electric trapping and shuttling potentials are generated. After that, the measurement scheme using the ion as a local field probe is presented. Then, the quantitative experimental results are discussed and compared to numerical simulations of the trapping fields.

\section{Experimental field generation}
\label{sec:realization}
We confine a single, laser cooled $^{40}\mathrm{Ca}^+$-ion in a linear Paul trap. The radial confinement is generated by a harmonic radio-frequency pseudo potential resulting from a $25~\mathrm{MHz}$ drive with an amplitude of about $350~\mathrm{V_{pp}}$. This results in radial trap frequencies around $3.5~\mathrm{MHz}$. For the axial confinement, 32 pairs of electrodes  are available. Each of these 64 segments can be individually biased to an voltage $V_i$, $|V_i|\leq 10~\mathrm{V}$. In the trap region the experiments were performed in, the segments are $250~\mathrm{\mu m}$ wide and separated by $30~\mathrm{\mu m}$ gaps (details can be found in~\cite{Schulz}). The voltage data preprocessed by a computer and transmitted to a home-built electronics device. This device houses an array of 16 serial input digital-to-analog converters (dac, $50~\mathrm{MHz}$ maximum clock frequency), each of which supplying four individual voltage lines, so that a data bus conveying 16 bit in parallel suffices to obtain the desired 64 outputs. The amplitude resolution of the converters is 16 bit with a measured noise level well below the last significant bit. The serial input design of the converters keeps the amount of data being transmitted in parallel at a minimum. This makes the design extendable to a large number of channels.

In this work, we restrict ourselves to the investigation of harmonic potentials. This is rectified by the fact that a resting, laser cooled ion only experiences the very minimum of the potential. Such potentials are generated by applying control voltages to the trap segments. Thereby, each specific voltage configuration results in an axial trap with a given trap frequency; the position of the potential minimum -- this is where the ion resides -- can be changed to shuttle the ion between different locations on the trap axis. Doing so, two requirements have to be met by the time dependent voltages: (i) The amplitudes must range within the limits given by the dac electronics, $\pm 10~\mathrm{V}$ in our case. (ii) Each segment's voltage is to change as continuous as possible to avoid unfeasibly high frequencies. All these requirements are fulfilled by the field simulation and voltage calculation techniques presented in sections~\ref{sec:app_fieldcalc} and \ref{sec:app_regul}, respectively. For the experiments presented here, it is advantageous to perform slow, adiabatic potential changes, and thus avoiding excess oscillations of the ion. For that reason, the update rate of digital to voltage converters was chosen with $\sim 3.3~\mathrm{ms}$, and a parallel port connection between computer and dac device was used. In an alternative operation modus of the trap supply device, the data transmission and updated rate is significantly sped up by using a field programmable gate array. Then we reach $4~\mathrm{\mu s}$ even when updating all 64 channels simultaneously. For the application presented here, it is sufficient to use a Doppler cooled probe ion. The probe's spatial extent can then be estimated from the ion temperature and the consequential extent of the motional wave function to be $\sim 60~\mathrm{nm}$. This value could be reduced to $\sim 5~\mathrm{nm}$ by sub-Doppler cooling and a tighter confinement of the ion.
\section{The local field probe measurement scheme}
\label{sec:spectroscopy}
To test both the numerical methods and the developed electronic devices for accuracy, we employ a single ion as a probe for the electrostatic potential. We exploit the fact that it is possible to exactly measure the trap frequency of the confining axial potential by spectroscopic means. Consider, for example, a voltage configuration $\left\{V_i\right\}^{(z_p)}$ to be tested for; this configuration is meant to result in a trap at position $z_p$ with frequency $\omega(z_p)$. Then we can use time dependent potentials to shuttle the ion from a loading position to $z_p$~\cite{Huber}, whereby the final voltage configuration must be $\left\{V_i\right\}^{(z_p)}$. The spectroscopically obtained trap frequency at $z_p$ can then be compared to the theoretically expected one, $\omega_\mathrm{sim}(z_p)$.

The trap frequency is determined by measuring the difference frequency of the carrier and the first motional red sideband (rsb) excitation of the ion in a resolved sideband regime. For that, we excite the quadrupole transition  $(4S_{1/2}, m_J=+1/2) \leftrightarrow (3D_{5/2}, m_J=+5/2)$ at $729~\mathrm{nm}$. As the carrier resonance frequency does not depend on the trap frequency\footnote{Nevertheless, the carrier frequency was monitored to exclude drifts in laser frequency or Zeeman splitting.}, it is sufficient to measure the rsb frequency at different locations $z_p$.
\begin{figure}
\resizebox{0.45\textwidth}{!}{%
  \includegraphics{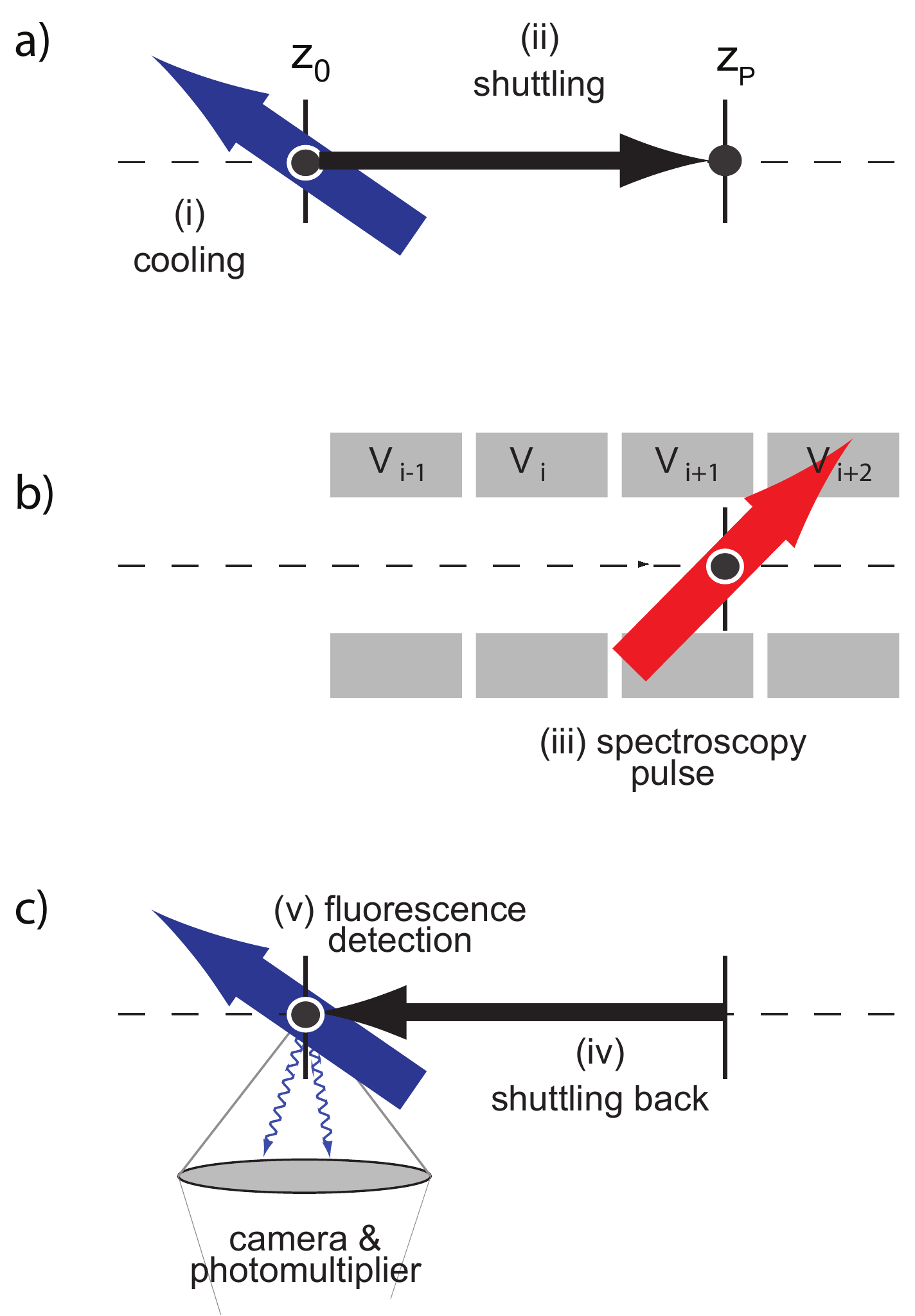}
}
\caption{Illustration of the measurement procedure: (i) cooling and preparatio of the ion at starting position $z_0$ and (ii) transport of the ion to the probing position $z_p$. A spectroscopy pulse with a certain detuning excites the ion there (iii). After shuttling the ion back (iv), the quantum state of the ion is read out (v). After many repetitions for different detunings an excitation spectrum of the ion at $z_p$ is obtained, from which the trap frequency $\omega(z_p)$ can be deduced.}
\label{fig:spectroscopy_scheme}       
\end{figure}
\par
The measurement scheme is illustrated in figure~\ref{fig:spectroscopy_scheme}. It consists of the following steps:
\par
(i) An initial voltage configuration is chosen to trap and cool the ion at the starting position $z_0$.
All lasers necessary for cooling, repumping, state preparation and detection are aligned to interact with the ion here. Additionally, $z_0$ is the position where fluorescence emitted by the ion can be detected by a photomultiplier tube and a camera.
\par
(ii) The ion is shuttled to the probing position  $z_p$. This is accomplished by applying voltage configurations resulting in a series of harmonic potentials whose minimum positions lead the ion from $z_0$ to $z_p$. For simplicity, these potentials were chosen such that their harmonic frequencies were almost constant and close to $\omega(z_p)$.
\par
(iii) Applying a spectroscopy pulse at the probing position $z_p$. Now, the voltages are exactly $\left\{V_i\right\}^{(z_p)}$.
Resting at $z_p$, the ion is exposed to a spectroscopy pulse of fixed duration ($100~\mathrm{\mu s}$).
The pulse is detuned from the (carrier) resonance by $\Delta f$.
This excites the ion with a probability $P$ into the upper state $\ket{D_{5/2}}$.
\par
(iv) Shuttle the ion back to $z_0$, inverting step (ii).
\par
(v) Having arrived back at the starting position, the state of the  ion is read out by illuminating it on the cooling transition. Whenever we detect a fluorescence level above a certain threshold, the ion is found in the ground state $\ket{S_{1/2}}$, while a low fluorescence level indicates that the ion has been excited to the state $\ket{D_{5/2}}$.
\par
The excitation probability for a specific detuning, $P(\Delta f)$, at the remote position $z_p$ is obtained by averaging over many repetitions of steps (i) to (v). By varying $\Delta f$, a spectrum of the quadrupole excitation at the remote position is obtained without moving any lasers or imaging optics but the spectroscopy laser used in step (iii). This is a significant advantage, because that way even such trap positions can be investigated that cannot be directly observed by the imaging devices or reached by all lasers.
\par
\begin{figure}
\resizebox{0.45\textwidth}{!}{%
  \includegraphics{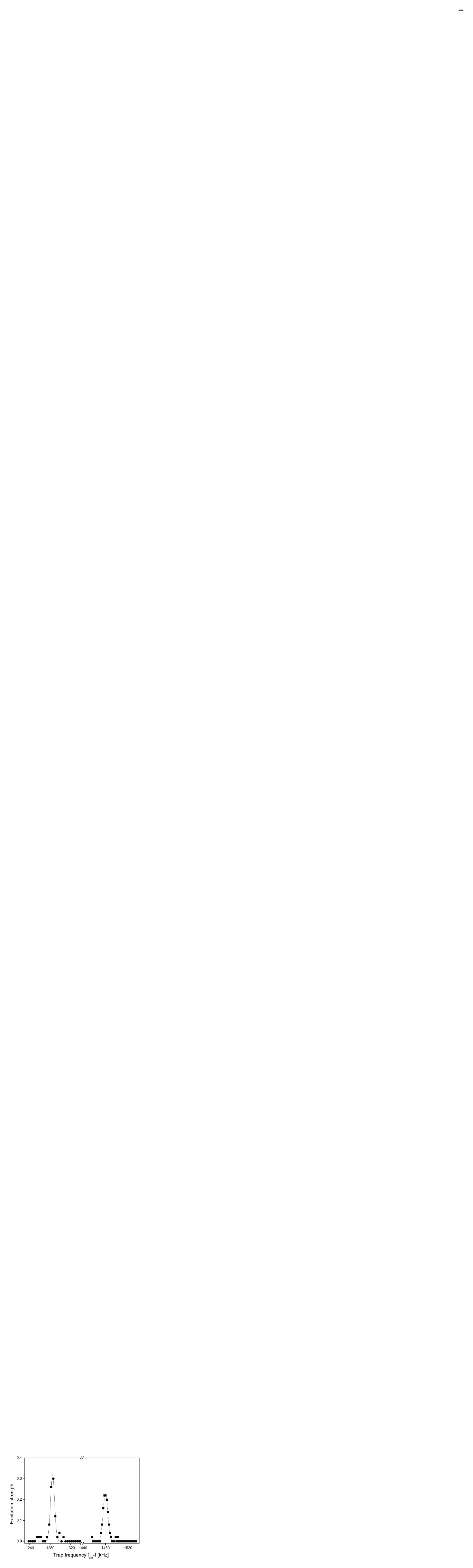}
}
\caption{Resonances of the first red sideband excitation on the $S_{1/2}\rightarrow D_{5/2}$ transition. The two peaks are measured at different trap positions along the trap axis. The resonance frequencies are given with respect to the carrier resonance $f_\mathrm{car}$, so that the trap frequency at the respective trap position can be read off.
The full width at half maximum of the peaks is $8~\mathrm{kHz}$ and determines the measurement accuracy.
The different peak heights stem from slightly different field intensities experienced by the ion.}
\label{fig:spectra}       
\end{figure}

The frequency difference between the red sideband and the carrier transition yields the trap frequency. Figure~\ref{fig:spectra} shows two rsb resonance peaks obtained at different positions within the trap volume.

\par
It is noticeable that each iteration cycle (i-v) including two ion transports results -- due to the binary nature of the projective readout -- in exactly one bit information about the spectrum. Therefore, thousands of transports, each relying on the calculated potentials, are performed for the determination of one frequency $\omega(z_p)$.
The transport, however, can be performed so fast ($\sim 100~\mathrm{\mu s}$) that its contribution to the overall experiment duration is secondary; this duration is still dominated by cooling and detection times ($\sim\mathrm{ms}$).
\section{Implementation and results}
To implement the measurement scheme, we first calculated voltage sets $\left\{V_i\right\}^{(z)}$, where $z$ covers the whole extent of the trap in steps of $5~\mu\mathrm{m}$ (For arbitrary positions, the voltages can be interpolated). Each set results in a certain, wanted frequency $\omega_\mathrm{sim}(z)$. That means that for each arbitrary position $z$ in the trap, there can be found a set of voltages resulting in a potential with its minimum at $z$ and with trap frequency $\omega_\mathrm{sim}(z)$. Then, in order to shuttle the ion, we simply subsequently apply the voltage configurations for $z=z_0\,...\,z_p$.
\par
The calculated voltages are tested with high axial resolution, i.e. in small steps of $z_p$, in two far distant regions of the trap. Doing this, both small local deviations should be detectable and the stability over the whole trap structure can be tested for.
To see a variation in $\omega(z_p)$ when increasing $z_p$, it is of advantage that small variations of the trap frequency around its means value occur. This is a reliable way to ensure that the ion in fact probes the remote position.
\par
\begin{figure}
\resizebox{0.48\textwidth}{!}{%
  \includegraphics{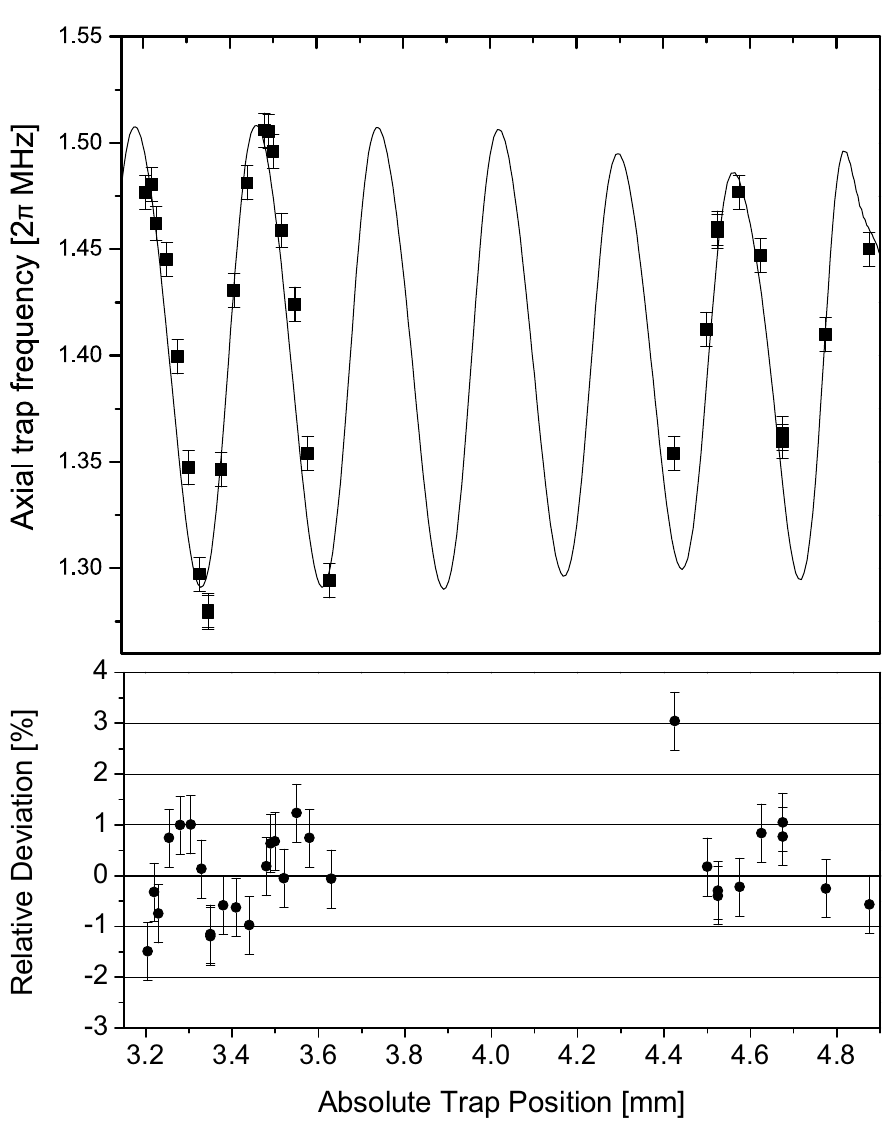}
}
\caption{(a) Harmonic trap frequency as a function of the position along the trap axis. The electrode voltages were calculated and applied such that the trap frequency shows a small oscillatory variation around a mean value of about $1.4~\mathrm{MHz}$ to see the ion proceed along the axis. The solid line shows the frequency of the wanted harmonic potential. The data points are the spectroscopically measured, real trap frequencies. The error bars show the uncertainty due to the finite resonance linewidth. (b) Relative deviation of the measured from the simulated frequency  $|\omega(z)-\omega_\mathrm{sim}(z)|/\omega_\mathrm{sim}(z)$.}
\label{fig:remote_results}       
\end{figure}
Figure~\ref{fig:remote_results} shows the expected, simulated trap frequencies and the measured ones. The data are in excellent agreement with the predicted frequencies. On both ends of the investigated trap structure, the predicted course of $\omega_\mathrm{sim}(z)$ is confirmed within the spectroscopic accuracy of $0.6\%$. The mean deviation of all measured data points is only $0.73\%$. Note that the solid line shown in figure~\ref{fig:remote_results} is based solely on geometric data from a technical drawing of the trap; there is no free parameter that was used to match the simulations with the measurement.
\section{Discussion and outlook}
\label{sec:conclusion}
The high accordance between the simulated and the real trap frequencies, as can be observed in the presented experiments, indicates the reliability of at least four independent contributions:
First, the numerical field simulation is accurate. This simulation is independent from a specific set of voltages, but relies on a model of the three dimensional trap design.

Second, the physical realization of this geometry is very good, i.e. the manufacturing and assembling process of the micro trap is so precise that it is not limiting the field accuracy. This aspect is coming more an more into the experimental focus, since the miniaturization of the trap layouts still proceeds. For example, a possible  axial misalignment $\Delta x$ of the two dc electrode wings with respect to each other would mainly result in an axial shift of the oscillations depicted in fig.~\ref{fig:remote_results}a by approximately $\Delta x/2$. Additionally, the amplitude of these oscillations decreases with increasing $\Delta x$. Indeed, the residuals (fig.~\ref{fig:remote_results}b) show a small oscillatory behavior. This oscillation signal is shifted by at most $\sim 15~\mathrm{\mu m}$ with respect to the trap frequency oscillation. Such an effect could stem from a small misalignment of the two trap layers (see ref.~\cite{Schulz}) or from a small offset of the theoretical and the real, axial zero point of position. We attribute the observed shift to an offset error, since in this case, the amplitude of the oscillation is not decreased, as it is the case in the measured data\footnote{The misalignment error can also be verified to be smaller than $\sim 10~\mathrm{\mu m}$ from microscope pictures of the trap.}. This shift is systematic and could be corrected for. Third, the calculation of the voltages leading to the wanted potentials is highly reliable. This is also qualitatively confirmed by the successful shuttling process itself. From our measurement we find that the generation of the voltages and their supply to the trap electrodes is indeed highly accurate. We also conclude that trapping fields from the applied voltages are not significantly perturbed by background fields from stray charges.

As the generation of trapping fields in the multi-segment trap was found to be accurate over the whole considered volume, it is possible to tailor potentials for special purposes. For quantum computing tasks, a harmonic potential of constant frequency is mostly required. Simulations show that it is possible to obtain fixed frequency potentials with a relative frequency deviation on the order of $10^{-4}$. The fast transport of quantum information in harmonic wells also requires a high degree of control over the trap voltages~\cite{Murphy}. The application of optimal control methods can help optimizing tailored time dependent potentials. Alternatively, one can feed-back information, gained from the measurement presented here, into the generation process to refine the results iteratively~\cite{Eble}.

However, the scope of possible applications of the described techniques is much wider than quantum computing tasks: Ions in time dependent potentials were proposed to be used as a testbed for quantum thermodynamic processes~\cite{Huber2} or for quantum simulation~\cite{Schuetzhold,Johanning}. The presented method is just a first proof of principle of using single atoms as ultra-sensitive probes. The measurement principle is not restricted to the investigation of the electric trapping fields. Only slight modifications allow for the detection of any electric or magnetic field with a nano-meter scaled quantum probe, as an alternative to cold atom sensors \cite{WILD06,KRUEGER07}. Then, a single atom field probe might be used as well for probing magnetic micro-structures with a relative accuracy better than 10$^{-3}$. Also, a study of decoherence and heating effects and their dependence on the ion-electrode separation or ion location is in reach.

Financial support from the DFG within the SFB/TRR-21, the German-Israel Science Foundation and by the European Commission within MICROTRAP, EMALI and SCALA is acknowledged.

\begin{appendix}
\section{Field calcuations}
\label{sec:app_fieldcalc}
To create a specific electric potential $\phi(z)$ on the trap axis, it is necessary to find the right set of voltages $\{V_i\}$ being applied to the trap electrodes labeled by $i=1,\ldots,N$. The potential generated by such a set of voltages is the linear superposition of the $N$ individual electrodes, whereby the contribution of each electrode $i$ is weighted by the applied voltage $V_i$.
After subdividing the axial position into a grid of $M$ points $z_j, j=1,\ldots,M$, we can write the overall potential at any $z_j$ by
\begin{align}\label{eq:matrix}
\phi(z_j)=\phi_j&=\sum_{i=1}^{N} A_{ij}\cdot V_i \notag
\end{align}
Here we introduced the electrode potential matrix $A_{ij}$. It describes the influence of the $i$-th electrode to the overall potential at $x_j$. Each row $i$ of $A$ can be seen as a position-dependent function, describing the potential generated by electrode $i$ (in units of $V_i$), when all other electrodes are set to zero voltage.
This quantity $A$ is independent from the specific voltage and is solely given by the trap geometry, i.e. the shape and size of the electrode and its distance from $z_j$, for instance.

Therewith, the potential generation can be logically divided into two parts: The matrix $A$ can be calculated independently from voltage constraints and independent from the desired potential. Second, for each desired potential $\vec\phi$, there has to be found a set of voltages $\vec v$ by inverting the matrix equation above. Tackling the first problem, one recognizes that modern segmented trap geometries can be realized in such a geometric complexity that conventional simulation techniques like the finite element method (FEM) fail. Instead, we obtained the potentials by solving the boundary element problem of the segmented trap design. Details can be found in~\cite{Greengard:1988,Nabors:1994,SingerRMP2009,Pozrikidis}.

\section{Calculation of the voltages}
\label{sec:app_regul}
In the following, we address the problem how to obtain a set of voltages $\{V_i\}$ that generates a given potential $\vec \phi$ when applied to the respective electrodes. The problem is formally solved by matrix inversion as $A^{-1} \vec \phi$. Several circumstances make this straight forward ansatz unfeasible: First, there is often no exact solution to the problem, because $\vec\phi$ is not an exactly realizable potential (note that in general, $M\gg N$ is possible). In this case, an approximate solution has to be found. Additionally, as a specific electrode does not effectively contribute to the potential at a far distant point, this electrode's voltage is ill-determined. These cases have to be treated adequately by the algorithm. We solved the inversion problem with a singular-value decomposition of the matrix $A$ to identify its critical, singular values. The real $N \times M$ matrix $A$ is decomposed into the product
\begin{equation}
A = USW^T,
\label{eq:}
\end{equation}
of the unitary matrices $U$ ($N\times N$) and $W$ ($M\times M$) and the diagonal $N \times M$ matrix $S$ with non-negative entries $s_k, k=1,\ldots,\min(M,N)$. This decomposition is part of many standard numerical libraries and can be performed for any input matrix $A$.
The inverse can then be written as
\begin{equation}
A^{-1}=WS^{-1}U^T.
\label{eq:}
\end{equation}
This step is numerically trivial, because the unitary matrices are simply transposed and the entries of $S^{-1}$ are given by $1/s_k$. At this point, the advantage of the decomposition becomes obvious, since small  values of $s_k$ indicate an (almost) singular, critical value. One way to overcome these singular values would be to cut off their diverging inverse values. Instead, the Tikhonov regularization~\cite{Tikhonov} method implies a more steady behavior as it makes the displacement $1/s_k \rightarrow s_k/(s_k^2+\alpha^2)$. The latter expression behaves like the original $1/s_k$ for large values $s_k\gg \alpha$, has its maximum at $s_k=\alpha$ and tends to zero for small, critical values $s_k\ll \alpha$. From this can be seen that the choice of the optimization parameter $\alpha$ is a compromise between exactness and boundedness of the results. For $\alpha=0$, the exact solution (if existent) is obtained, whereas large values of $\alpha$ guarantee small inverse values and thus bounded voltage results. Therefore, we label the regularized quantities with index $\alpha$. The approximate solution $\vec v_\alpha = (V_1, \ldots, V_N)$ is
\begin{equation}
\vec v_\alpha = WS^{-1}_\alpha U^T \vec \phi,
\label{eq:phi1}
\end{equation}
with $S^{-1}_\alpha$ being the regularized matrix with entries $s_k/(s_k^2+\alpha^2)$.
\par
Before the problem of finding an optimal $\alpha$ is addressed, another constraint regarding time dependent voltages, i.e. series of voltage configurations, has to be accounted for. While moving the ion by one step, i.e. from a position $z_p$ to $z_p'$, the control voltage should vary as less as possible. This is achieved by extending eqn.~(\ref{eq:phi1}),
\begin{equation}
\vec v_\alpha' = \vec v_\alpha + WD_\alpha W^T\vec{v},
\label{eq:phi2}
\end{equation}
where $\vec{v}$ represents any previous voltage set, providing trapping at $z_p$. The second term in eqn.~(\ref{eq:phi2}) contains a diagonal matrix $D$ with entries $d_k= \alpha^2/(s_k^2+\alpha^2)$. $d_k$ tends to zero for $s_k \gg \alpha$, so that uncritical voltages are affected only little by the second term. For all critical voltages indicated by a value $s_k<<\alpha$, however, the first term in eqn.~(\ref{eq:phi2}) vanishes due to the regularization replacement and what remains is the contribution from $\vec{v_0}$, since then, $d_k \approx 1$. Here, the choice of $\alpha$ determines, how strong the algorithm tries to generate similar voltages in a (time) series of voltage sets.

The algorithm described above minimizes $ ||A\vec{v}'_\alpha-\vec{\phi}||^2+ \alpha||\vec{v}'_\alpha-\vec{v}||^2$ with respect to the Euclidian norm. That is, the potential $\vec \phi$ is reproduced as good as possible under the constraint that solutions similar to the previous one are preferred. What remains is to find the proper value of $\alpha$. Hereby, one has to make a tradeoff between the boundedness of the voltages and their continuity. Under practical circumstances requiring $|V_i|\leq V_\mathrm{max}$ for some maximal voltage $V_\mathrm{max}$, $\alpha$ can be iteratively increased to fulfill this constraint on the one hand, and to obtain as continuous voltage sets as possible, on the other hand.
\end{appendix}
%

\begin{thebibliography}{}
%
%
\bibitem{Benhelm}
J. Benhelm et al., Nature Physics \textbf{4}, 463 (2008).
\bibitem{Haeffner}
H. H\"affner et al., Nature \textbf{438}, 643 (2005).
\bibitem{Leibfried}
D. Leibfried et al., Nature \textbf{438}, 639 (2005).
\bibitem{Chiaverini}
J. Chiaverini et al., Nature \textbf{432}, 602 (2004).
\bibitem{Leibfried2}
D. Leibfried et al, Phys. Rev. A \textbf{76}, 032324 (2007).
\bibitem{Kielpinski}
D. Kielpinski, C. Monroe  and  D. J. Wineland,  Nature \textbf{417}, 709 (2002).
\bibitem{Schulz2006}
 S. Schulz et al., Fortschr. Phys. \textbf{54}, No. 8-10, 648 (2006).
\bibitem{Greengard:1988}
L. Greengard and V. Rokhlin, in: Vortex Methods, 121, (Springer, Berlin 1988).
\bibitem{Nabors:1994}
K. Nabors et al., Sci. Comput. \textbf{16}, 713 (1994).
\bibitem{Eble}
J. Eble et al., accepted for publication in J. Opt. Soc. Am. B (2010), arXiv:0912.2527.
\bibitem{SingerRMP2009}
K. Singer et al., arXiv:0912.0196.
\bibitem{Schulz}
S. Schulz et al.,  New J. Phys. \textbf{10}, 045007 (2008).
\bibitem{Huber}
G. Huber et al., New J. Phys. \textbf{10}, 013004 (2008).
\bibitem{Huber2}
G. Huber et al.,  Phys. Rev. Lett. \textbf{101}, 070403 (2008).
\bibitem{Schuetzhold}
R Sch\"utzhold et al., Phys. Rev. Lett. \textbf{99}, 201301 (2007).
\bibitem{Johanning}
M. Johanning et al.,  J. Phys. B: At. Mol. Opt. Phys. \textbf{42}, 154009 (2009).
\bibitem{Murphy}
M. Murphy et al., Phys. Rev. A \textbf{79}, 020301(R) (2009).
\bibitem{Wunderlich}
H. Wunderlich et. al., Phys. Rev. A \textbf{79}, 052324 (2009).
\bibitem{Pozrikidis}
C. Pozrikidis, A Practical Guide to Boundary Element Methods with the software library BEMLIB, Chapman and Hall/CRC, Boca Raton, FL, USA, (2002).
\bibitem{Tikhonov}
A. N. Tikhonov and V. A. Arsenin, Solution of Ill-
posed Problems, Winston and Sons, Washington, USA (1977).

\bibitem{KRUEGER07}
P. Kr\"uger, et al., Phys. Rev. A \textbf{76}, 063621 (2007).
\bibitem{WILD06} 
S. Wildermuth et al., Appl. Phys. Lett. \textbf{88}, 264103 (2006).



\end{thebibliography}
%

\end{document}